# PeopleXploit
*A hybrid tool to collect public data*

Arjun Anand. V
*Information Technology*
*SSN College of Engineering*
*Kalavakkam – 603 110, India*
arjunahalya99@gmail.com

Buvanasri A K
*Information Technology*
*SSN College of Engineering*
*Kalavakkam – 603 110, India*
ananthubuvanasri@gmail.com

Meenakshi R
*Information Technology*
*SSN College of Engineering*
*Kalavakkam – 603 110, India*
meenakshiramesh2308@gmail.com

Dr. Karthika S
*Information Technology*
*SSN College of Engineering*
*Kalavakkam – 603 110, India*
skarthika@ssn.edu.in

Ashok Kumar Mohan
*TIFAC-CORE in Cyber Security,*
*Amrita School of Engineering,*
*Coimbatore*
*Amrita Vishwa Vidyapeetham,*
*India*
m_ashokkumar@cb.amrita.edu

*Abstract*—This paper introduces the concept of Open Source Intelligence (OSINT) as an important application in intelligent profiling of individuals. With a variety of tools available, significant data shall be obtained on an individual as a consequence of analyzing his/her internet presence but all of this comes at the cost of low relevance. To increase the relevance score in profiling, PeopleXploit is being introduced. PeopleXploit is a hybrid tool which helps in collecting the publicly available information that is reliable and relevant to the given input. This tool is used to track and trace the given target with their digital footprints like Name, Email, Phone Number, User IDs etc. and the tool will scan & search other associated data from public available records from the internet and create a summary report against the target. PeopleXploit profiles a person using authorship analysis and finds the best matching guess. Also, the type of analysis performed (professional/ matrimonial/criminal entity) varies with the requirement of the user.

*Keywords*—OSINT, SOCMINT, cyber security, profiling, analysis, social media, intelligence.

## I. Introduction

Technology has become an integral part of today's fast paced world. Right from coffee machines to heat seeking missiles, technology has influenced every aspect of daily routine life. Long distance communication, about 20 years back, was mainly based on physical devices like telephones to perform voice calls and meeting face to face meant the person actually had to go all the distance. Nowadays, a single application can be used to send and receive digital messages, make voice calls with premium quality and even perform video calls where face to face communication has been brought down to simple one or two button procedures. Twenty-five years ago, people had to walk to the bank and interact with a bank employee or bank teller to deposit a paycheck or withdraw cash. But banks have realised that they are not just selling a service to their customers, but an experience. So, they have made online banking a convenient way for customers to make transactions from their comfortable places or on-the-go with mobile banking. The increase in the importance of a virtual interface for different basic services has been focused mainly on customer satisfaction over blunt profit margins. Infact, many people appraise online banking to be the most functional lifestyle innovations. Take a step forward, 69% of European millennials manage their finances through mobile devices [1].The most important idea that has affected the evolution of technology to this magnitude is the concept of internetworking of devices. Advancement in this area has enabled people to become faceless or masked in the virtual space. This has both positive and negative consequences. While privacy of the common layman has been prioritized and emphasized, unmasking of cyber criminals has also become a major problem. The virtual space provided by this network has perpetuated the rise of a lawless dimension in which any action can be performed to cause as much havoc as a terrorist attack and also be as beneficial as blockchain in securing intellectual properties.

With an ever increasing amount of people getting connected to the internet with each passing day, the security threats that cause massive harm are also increasing. Cyber security is referred to as information technology security. It may also be referred to as the body of technologies, processes, and practices designed to protect networks, devices, programs and data from damage, attack or unauthorized access. Merriam Webster dictionary defines it as "Measures taken to protect a computer or computer system (as on the Internet) against unauthorized access or attack" [2]. As defined by the International Telecommunications Union (ITU) cyber security is "The collection of tools, policies, security concepts, security safeguards, guidelines, risk management approaches, actions, training, best practices, assurance and technologies that can be used to protect the cyber environment and organization and user's assets" [3]. Any organization or basically any user can have connected computing devices, infrastructure, telecommunications systems and probably the entire transmitted/stored data in a digital habitat. Cyber Security seeks to tackle the risks and threats that persist in such a digital habitat and ensure that the basic security objectives(Confidentiality, Integrity and Availability) are satisfied through and through [4].

As the technological revolution advances towards the next level, the amount of generated information also increases exponentially. Information is everywhere. Data is the new "wheel". Although mining of data without user consent is punishable by law, the value of user-shared information is undeniable. Open source Intelligence (OSINT) is the "act of finding information". It is "intelligence that is produced from publicly available information and is collected, exploited, and disseminated in a timely manner to an appropriate audience for the purpose of addressing a specific intelligence requirement" [5]. Almost every individual in this world performs OSINT on a regular basis. How many times does one use a search engine looking up for answers? That is OSINT in a nutshell. During this process, information is searched across publicly available sources, although the term isn't strictly limited to the internet. Online sources such as social media profiles, blog entries or media viz., videos, images are primarily used for this but more professional sources such as newspapers or business reports are also combed through. Though the information acquired is free and legal, it isn't classified. Such a harvest consists of a huge, relatively complex collection of data which first needs to be structured and correlated.

'Social Media' is one of the most prominent ways in which the internet is increasingly being used to participate, to create and to share information about people and their associated friends, their likes and dislikes, expressions, thoughts and transactions. Society as a complete unit often creates and adopts a new method to communicate and organise through every step of technological evolution, hence it is vital for public infrastructures like national security to evolve as well [6]. SOCMINT (Social Media Intelligence) extends over a wide range of applications, methodologies and capabilities available through the collective use of social media data. It is an extended division of OSINT and has been defined as 'information that is publically available and can be lawfully obtained by request, purchase or observation'. SOCMINT requires very precise considerations of credibility and understanding [7].

## Literature Survey

One of the famous microblogging platforms which has billions of users to propagate billions of personal or professional posts, stories on a circadian substructure is twitter [8]. Social Bearing is an analytical tool that allows the users to do a search on keywords, geolocation, hashtag or people on Twitter. It enables its user to understand their audience through custom reports on audience sentiments, source and behavior data. The reach of the tool is extensive, it essentially sums up the number of followers of each individual tweeting and retweeting. Tweet by geolocation can also be done with the map in the tool. Just like Google Analytics, the deeper the green indicates the increased activity. The deliverables of the tool are available as a downloadable csv file. But the entire research of the tool is based only on twitter. PeopleXploit focuses on enhancing its output by not just on twitter, but also on all social media and the internet.

## Related Work

For emphasizing the value of OSINT, let us consider the case of 35-year-old Shahin Gheiybe, a "Most Wanted" Dutch fugitive. On July 5, 2011, Shahin Gheiybe was sentenced to a term of incarceration for thirteen years on two counts of attempted murder. He absconded from the country on October 6, 2011, and has been on the run ever since. The Korps Nationale Politie(The National Police Corps.)have since tried to track and bring the fugitive to justice but all their attempts have been in vain. Shahin Gheiybe actively posted on Instagram, with over 170 photos and videos, where most of them were taken in Iran. Shahin took to his Instagram handle to convince people that followed him that he was just another normal man who had made a few mistakes and was in reality, innocent. Although he never actually tagged any geolocation in his posts, the content that he was posting enabled Bellingcat - an online community that practices investigative journalism and open source intelligence - to track him down and trace his trail using very subtle hints like the foliage at the background of a video and the number plate of a car in one of the posts [9]. This is an example of how OSINT works in both ways: it can be used for both benefit and downfall of people.

With the increase in the use of digital data, threats like phishing have also been increasing. Ever wondered why you are getting free cryptocurrency from the founder of Tesla? Phishing is a kind of social engineering attack mostly used to plunder user-sensitive information, which may include login credentials and credit card numbers. It transpires when an attacker, camouflaging as a trusted and known object, dupes a victim into opening an email, instant message, or text message. The receiver is then misled into clicking a harmful link, which can ultimately lead to infection by malware, the suspension of the system as part of a ransomware attack or the breach of sensitive information. OSINT has been a major influence in preventing such masked menaces.

There have been numerous reports of criminals escaping the whips of justice by taking advantage of the extensive time taken for individual identification by law enforcement agencies. Tamil Nadu sits at the top 3 states in India in the number of crimes per lakh people with a score of 294/1, 00,000. Today, open source tools are available to collect data from publicly available sources to be used in an intelligence context. These tools simplify the process of searching through large volumes of data. Study of these publicly available data can help us analyze social interactive patterns, interrelated clues and important hidden relations between various objects, scenarios, people, etc.

The United States Intelligence Community has been compelled to substantially increase the use of Open Source Intelligence (OSINT). OSINT contributes to intelligence operations and increases the efficiency of other methodologies deployed for intelligence purposes by providing a precise understanding as to what kind of data is available readily and what kind of data requires dedicated focus and extraordinary measures to extract. By using OSINT techniques, Intelligence agencies can reduce their dependencies on aggressive extraction strategies like espionage or stealing and reduce the probability of miscalculations and aggravation of unwanted tensions between countries. OSINT methodologies also reduce the costs of implementing other intelligence operations by reducing ambiguity in data to a great extent [10].

In this paper, the authors address the above mentioned problems. As it is not possible to collect data regarding the target by a single tool, we have come up with a stack of tools to do the analysis and background checks using a target's digital presence. The stack of tools are combined into a hybrid tool.

The following are the objectives of the tool:

1. To identify possible individual wrongdoers from socially shared information.

2. To perform background checks on the employees of a company.

3. To be able to perform bride/bridegroom background checks on matrimonial sites.

4. To examine real-time and pre-existing/ historical data sets for profiling and develop a real time Hybrid tool (tool stack) for effectively identifying targets.

## II. METHODOLOGY

The proposed tool works for any input given in the form of image or text. An OSINT search can be performed on a particular individual and a report can be generated by tracking their digital presence. For example, if an input, like the email-ID of the wrongdoer is given, it can be reverse searched all over the internet to check for associated accounts on social media, blogs, etc. and for any openly available, relevant information. Using this process, we will be able to construct a digital profile for the target individual. Undertaking this particular application of OSINT, the authors had performed some real time tasks as listed in table

TABLE I. TOOLS USED FOR REALTIME OSINT TASKS

| INPUT | DESCRIPTION | TOOLS |
|---|---|---|
| Social media handle | Fake account identification | Webmii, Vivial, Upolos |
| Social media handle | Profile Analysis | Webmii, Rapportive, Vivial, Social Bearing, Tinfoleak |
| Name | Profile Analysis | Webmii, Maltego, Rapportive, Pipl |
| Mobile Number | Profile Analysis | Webmii, Maltego, B-Mobile |
| Email | OSINT Analysis | Searchbug, verify-email, WhatBreach |

The list of tools used for the above investigations are open source and most of them are web based. Although, the greatest benefit of these tools also acts as its weakness. The relevance of the data generated through these tools is not 100% accurate. Also, the data generated by these tools is derived from region specific databases. Hence, the information returned need not be relevant to the scope of the target. To tackle this problem of accuracy and relevance, a hybrid system that aggregates the functionalities of a static set of OSINT tools is necessary. This master tool can interact, feed data provided by the user and retrieve the output dataset from each tool. It can also aggregate this output dataset, filter relevant information and give a readable/ downloadable report to the user.

Figure 1 represents the architecture of PeopleXploit:

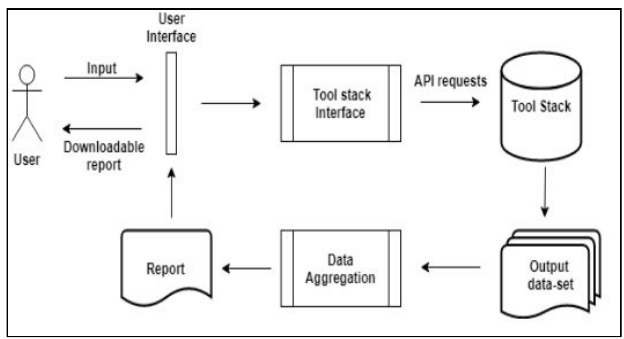

*Fig.1 Architecture of PeopleXploit*

The user enters an input data file in one of the following formats -
- Name - Person, Product, Organization, Institution, etc.
- Email - Gmail, Yahoo, Outlook, iCloud, Protonmail, etc.
- Phone Number - Airtel, BSNL, Jio, Vodafone, Idea and any foreign mobile dealer as well.
- Image - JPEG, PNG, GIF, TIFF, etc.

This data file will be forwarded to the Tool Stack Interface where it is populated as input to all the corresponding OSINT tools available in the Tool Stack. The Tool Stack then delivers an output dataset which comprises the data output of each individual tool. This dataset is further filtered and aggregated to improve relevance and precision of the data with accordance to the user's requirements. Finally, the collected data is returned to the user interface where it is submitted to the user in a downloadable format.

The authors were able to identify a certain set of tools that provide the best results. Using these tools, the above mentioned Tool Stack has been designed.
Some of the tools used in the Tool Stack are as follows:

**WebMii**:

WebMii is a people search engine.

- It centralizes public information from other websites.
- For each search, it shows the sources of the presented information.
- WebMii provides a vision of what's online about someone.
- WebMii shows the PeopleRank of a person: the PeopleRank is the online visibility score for someone.
- It is calculated by WebMii according to anyone's visibility on the web.
- This score can be used to evaluate and scope the amount of public information available about a given person.

**Upolos:**

Upolos is a social media tool which is used to view Instagram profiles via the browser.

- It also displays the posts that the target has put on instagram if the profile is public.
- This tool is used to preserve the anonymity of the user as it doesn't require an account to view profiles.

**Rapportive:**

Rapportive is a LinkedIn sales navigator. It is a chrome extension or in other words, is a plug-in for Gmail. Upon hovering over any email address in any message, Rapportive lets the user to

- See data from LinkedIn profiles of his/her Gmail contacts and allow them to be used when they require.
- Disclose icebreakers, such as connections that they share, their interests and experiences to stay informed or to help establish connections.

**Tinfoleak:**

Tinfoleak is a Twitter analysis tool. It produces a html report that contains the following information:

- Basic information that includes Twitter ID, URL, Screen Name, Location, Timezone, etc
- Client's source applications, associated social networks and topics used.
- Detailed information of all hashtags and mentions in tweets that can be viewed separately.
- Directories of images and videos respectively.
- Geolocation coordinates and place to generate a tracking map of locations visited.
- Show tweets by the user on Google Earth!

**StalkScan:**

Stalkscan is a free online service that enables to look up any Facebook user's public information using Facebook's own APIs**.** Given the URL of the target's Facebook, the following information can be collected:

- Profile: Public profile information including posts, pictures, groups, videos, past and future events, apps, and games.
- People: Friends, family, friends of friends, classmates, co-workers and locals.
- Interests: Pages, music, movies, books, political parties, religion, and places.
- Tags: Videos, pictures and posts the user tagged, or was tagged in.
- Comments: Comments the user posted on Facebook.
- Likes: Pictures, videos, or posts the user liked.
- Places: Places the user visited including bars, restaurants, stores, outdoors, hotels and theaters.

**Social Searcher:**

Social Searcher is a Social Media Search Engine. It provides a real-time search from 11 sources.

- Allows users to search for content in social networks in real-time and provides extensive analytics data.
- Users can search or look for content without logging in for publicly posted information on Reddit, Twitter, Facebook, Google+, Flickr, Instagram, Tumblr, YouTube, Vimeo, Dailymotion, and all Web generally.
- The analytics include mentions, users (in the same name) and sentiment.
- It also provides detailed information about the posts based on several criteria viz., time(hour, week day), social network; links and hashtags related to the target.
- Users can also save their searches and set up email alerts.

### III. Result

An OSINT search can be performed on any entity (object, individual) whose information is available in public resources such as the Internet. The objective of this paper is to examine real-time and pre-existing/ historical open source datasets for digital profiling. The presented case studies depicts the output format of PeopleXploit

**Objective 1**: To identify possible individual wrongdoers from socially shared information:

CASE STUDY:

ShahinGheiybe

**Bio**
Full Name: ShahinGheiybe
Date of birth: 01- 08- 1983
Place of birth: Tehran (Iran)
Sex: Male

**Physical characteristics:**

Height: 1, 88 m.
Hair colour: Black (short, wavy/curly)
Eye colour: Brown/ black

**Other Specifications:**
• Dimple in the chin
• Birthmark on left cheek near mouth

**Instagram handle:** @shahin.mzr
**Images:**

Refer Figure 2. For ShahinGheiybe's pictures

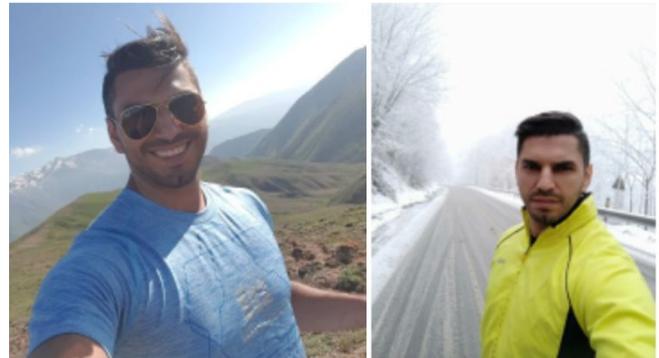

*Fig 2. Images of ShahinGheiybe from Instagram*

**Video:**
The following video was used by police to track his latest updated location by identifying hidden connections:
Click here to watch the video .

**Criminal Record -**
 5 July 2011, two attempted murders
• In 's-Hertogenbosch, on 6 July 2009, he tried to murder the two individuals with whom he had an appointment by shooting at them.
• He was sentenced to a term of imprisonment of thirteen years.
• Gheiybe managed to escape from prison on 6 October 2011 and has been on the run ever since.
• The police are looking for ShahinGheiybe .

The above case study depicts how the application will handle a criminal profile request [11][12].

**Objective 2:** To perform background checks on the employees of a company:

CASE STUDY:

Dr. Raja Reddy

**Bio**
Full Name: DabbalaRajagopal "Raj" Reddy
Date of birth: 13 June 1937
Sex: Male

**Education**
- Bachelor's Degree: Civil Engineering from College

of Engineering (1958)

- MTech: University of New South Wales, Australia (1960).

- Ph.D Degree in Computer Science from Stanford University (1966)

**Family**
- Parents- Sreenivasulu Reddy and Pitchamma
- Brothers: 7
- Sister: 1
- Daughter: 2

**Contact Details**
MozaBint Nasser University Professor, RI / CS
Office: 5327 Wean Hall
Phone: (412) 268-2597

**Career**
- MozaBint Nasser University Professor of Computer Science and Roboticsin the School of Computer Science at Carnegie Mellon University.
- For a brief period of time (1960-63), he worked as an Applied Science Representative for IBM Corp. in Australia.
- Dr.Raja Reddy was an Assistant Professor of Computer Science at Stanford from 1966-69.
- Dr.Raja Reddy joined as an associate professor in the Carnegie Mellon University in 1969.

**Awards**
- The Legion of Honor by FrenchPresidentFrançois Mitterrand in 1984.
- Padma Bhushan by Indian President Kocheril Raman Narayanan in 2001.
- The ACM Turing Award in 1994
- Okawa Prize in 2004
- In 2005, the Honda Prize.
- In 2006, theVannevar Bush Award.
- The doctorates (Doctor HonorisCausa) from SV University, Universite Henri-Poincare, University of New South Wales, University of Massachusetts, University of Warwick, Anna University, IIIT (Allahabad), Andhra University, Jawaharlal Nehru Technological University, IIT Kharagpur and Hong Kong University of Science and Technology.

**Research**
- Dr. Raja Reddy's field of interests are artificial intelligence and the study of human-computer interaction.
- His current research interests are as follows:
- Service of Society technology
- Guardian Angels and Cognition Amplifiers.

**Honours**
- His professional honors are as follows:
- Fellow of the Institute of Electrical and Electronics Engineers
- President of the American Association for Artificial Intelligence from 1987 to 1989.
- In 2011, Dr.Raja Reddy was invited into inaugural IEEE Intelligent Systems' AI's Hall of Fame.

The above case study depicts how the application will handle an employee profile request [13][14][15][16][17].

**Objective 3:** To be able to perform bride/bridegroom background checks on matrimonial sites:

CASE STUDY:

Harry Styles

**Bio**
- Birth Name: Harry Edward Styles
- Nickname: Hazza, Haz, Curly, Harold
- Profession: Singer, Songwriter, Actor

**Physical Stats & more**
- Height: 1.78 m
- Weight: 70 kg
- Eye Color: Hazel Green
- Hair Color: Dark Brown

**Personal Life**
- DoB: 1 February 1994
- Age (as of 2020): 26 years
- Birth place: Redditch, England
- Zodiac sign/ Sun sign: Aquaris
- Nationality: English
- Hometown: Redditch, English
- School: Holmes Chapel Comprehensive School, England
- Debut Album- Up All Night (2011)
- Family: Father- Desmond Styles
- Mother- Anne Twist
- Sister- Gemma Styles
- Brother(step)- Mike
- Religion: Roman Catholiscm
- Ethnicity: English
- Hobbies: Computer Gaming, Movies, Badminton and Tennis

**Favourite Things**
- Favourite Beverages: Apple Juice
- Favourite Computer Game: FIFA

- Favourite Animal: Turtle
- Favourite Food: Sweet Corn, Tacos
- Favorite Colours: Orange, Blue

**Partners and More**
- Marital Status: Unmarried
- GirlFriends: Caroline Flack (2011-2012)
    - Taylor Swift (2012-2013)
    - Kendall Jenner (2013-2014; 2015-2016)
    - Erin Foster (2014)

- o Georgia Fowler (2015)
- o Pandora Lennard (2016)
- Children: None
- NetWorth**: 23$Mil.

**Other habits**
- Drinking: Yes
- Smoking: Yes

The above case study depicts how the application will handle a bridegroom profile request [18][19].

Table II shows the type of input that can be given to each tool:-

TABLE II. INPUT THAT CAN BE GIVEN TO EACH TOOL

| Input type / Tool | Domain | Email address | Facebook handle | Instagram handle | Keyword(s) | Mobile number | Twitter handle |
|---|---|---|---|---|---|---|---|
| Bmobile | | | | | | ✓ | |
| Maltego | ✓ | ✓ | ✓ | ✓ | ✓ | ✓ | ✓ |
| Pipl | | ✓ | | | ✓ | | |
| Rapportive | | ✓ | | | | | |
| Searchbug | | ✓ | | | | | |
| Social Bearing | | | | | | | ✓ |
| Social Buzz | | | | | | | ✓ |
| Stalkscan | | | ✓ | | | | |
| Tinfoleak | | | | | | | ✓ |
| Upolos | | | | ✓ | | | |
| Verify email | | ✓ | | | | | |
| Vivial | ✓ | | | | | | |
| Webmi | | | | | ✓ | | |
| WhatBreach | | ✓ | | | | | |

### PROS AND CONS

1. PeopleXploit aims to act as a search-engine that only needs a name/ email address/ phone number as input that triggers a hunt for relevant information that reaches up to the deepest corners of the internet.
2. Searching for information about a person or a company can be a time consuming activity. PeopleXploit aims to collect information about the target in a minimal time period.
3. The processed information is made available in a structured, easy to understand report to the officers in charge.
4. It is possible to gather target's information such as phone numbers, social media handles from inputs such as email addresses and vice versa.

One of the disadvantages of PeopleXploit is that the filtration of noise can be difficult when the input is generalized. When a name is given as an input, the task becomes tedious as there would be a number of online profiles found with the same name. Hence, it requires a large amount of analytical work to distinguish the valid information.

## IV. CONCLUSION

The abundance of information has made our lives both difficult and easier for OSINT. It is easier because there is a wide range of channels for communication and difficult because of the proportion of junk or misleading information [20]. The mere magnitude of data broadens OSINT's objectives from not just fetching and processing digital data, but to also developing, verifying and understanding what constitutes irrelevant content and what doesn't. Though efforts are underway to increase the use of OSINT, there are many obstacles faced by the authors. One of the obstacles is the lack of tools helping to manage the volume of available data and ascertain its credibility. Another obstacle is the lack of relevance in the data collected by each tool.

Whenever some user posts on social media networks, knowingly or unknowingly their activity is registered in countless online repositories. This exposes parts of the data to be publicly available; as a consequence of this a forensic analyst can reveal past activities, reconstruct a biased timeline and recover deleted data of the suspect [21]. As reported above, the value of analyzing the online presence of an individual can reveal a lot about his/her true nature and it is up to the user to decide whether to use this information in a constructive way that promotes justice or to use it to monitor the opinion of the general public.

The authors have analyzed 15 tools and 3 case studies has been presented. In the future, advanced tools like Maltego and Shodan can be added to the tool stack of PeopleXploit. Maltego is a tool that enables us to actively gather information. Among the most popular and practical cyber intelligence tools, Maltego is a very practical information-gathering program that can be found built-in in Kali Linux. Shodan is a search engine that lets the user find specific types of computers (webcams, routers, servers, etc.) connected to the internet using a variety of filters. In the future, PeopleXploit can cater the needs of the users with additional features.